\documentclass[aps,pre,reprint,superscriptaddress,showpacs,preprintnumbers,amsmath,amssymb,longbibliography,nofootinbib]{revtex4-2}
\usepackage[dvipdfmx]{graphicx}
\usepackage{dcolumn}   
\usepackage{bm}        
\usepackage{amssymb, amsmath, amsfonts}   
\usepackage{lipsum}        
\usepackage{color}
\DeclareGraphicsExtensions{{.pdf}}
\begin{document}
\newcommand{\noter}[1]{{\color{red}{#1}}}
\newcommand{\noteb}[1]{{\color{blue}{#1}}}
\newcommand{\field}{\left( \boldsymbol{r}\right)}
\newcommand{\paren}[1]{\left({#1}\right)}
\newcommand{\vect}[1]{\boldsymbol{#1}}
\newcommand{\uvect}[1]{\tilde{\boldsymbol{#1}}}
\newcommand{\vdot}[1]{\dot{\boldsymbol{#1}}}
\newcommand{\vder}{\boldsymbol{\nabla}}
\widetext
\title{
Shear-induced Criticality in Glasses Shares Qualitative Similarities with the Gardner Phase
}
\author{Norihiro Oyama}
\email{oyamanorihiro0215@gmail.com}
\affiliation{Graduate School of Arts and Sciences, The University of Tokyo, Tokyo 153-8902, Japan}
\affiliation{Mathematics for Advanced Materials-OIL, AIST, Sendai 980-8577, Japan}
\author{Hideyuki Mizuno}
\affiliation{Graduate School of Arts and Sciences, The University of Tokyo, Tokyo 153-8902, Japan}
\author{Atsushi Ikeda}
\affiliation{Graduate School of Arts and Sciences, The University of
Tokyo, Tokyo 153-8902, Japan}
\affiliation{Research Center for Complex Systems Biology, Universal Biology Institute, University of Tokyo, Komaba, Tokyo 153-8902, Japan}

\date{\today}
\begin{abstract}
Although glass phases are found in various soft matter systems ubiquitously, we are still far from a complete understanding of them.
The concept of marginal stability predicted by infinite-dimensional mean-field theories is drawing attention as a candidate for a universal and distinguishing unique features of glasses.
While among theoretical predictions, the non-Debye scaling has indeed been observed universally over various classes of glasses, and the Gardner phase is found only in hard sphere systems if we restrict ourselves to the physical dimensions.
In this work, we numerically demonstrate that plastic events observed in two-dimensional Lennard-Jones glasses under quasistatic shear exhibit statistical properties that are qualitatively very similar to the Gardner phase.
\end{abstract}
\maketitle
%
\section{Introduction}
Glass transitions are observed ubiquitously in various soft matter systems \cite{Berthier2011,Berthier2011a,Tanaka2019}, and much effort has been dedicated to the exploration of the nontrivial features associated with this phenomenon over the past few decades.
However, our knowledge of the glass transition and resulting glass phases is still far from complete, being regarded as \emph{the deepest and most interesting unsolved problem in solid state theory}~\cite{Anderson1995}.
Recently a concept called the marginal stability has received much attention as a potential candidate for a universal and distinguishing unique feature of glass phases~\cite{Muller2015}: this is a characteristic of glass systems in which they susceptibly exhibit plastic deformations against even an infinitesimal deformation.
In particular, mean-field replica theories in infinite dimensions predict salient features associated with the marginal stability, such as the so-called non-Debye scaling~\cite{Franz2015} or the Gardner transition~\cite{Charbonneau2014,Charbonneau2015,Biroli2016}.

The first theoretical prediction, the non-Debye scaling, is related to the low frequency limit of the vibrational density of states (VDoS).
Though a complete description of the VDoS of crystals was provided by Debye's law~\cite{Kittel1996}, the counterpart for glasses is still missing.
According to the mean-field replica theory~\cite{Franz2015}, in the low frequency limit, glasses have peculiar extra modes in addition to the standard phonon modes that Debye's law describes.
These extra modes are considered to correspond to elementary processes of plasticity, or shear transformation zones (STZs)~\cite{Maloney2004}, and obey a different power-law scaling from that for Debye's law.
Considering that a larger system tends to possess vibrational modes with smaller frequencies,  
this non-Debye scaling indicates that thermodynamically large glass systems experience plastic events even under infinitesimal perturbations.
{Indeed, numerical results have verified that the average amount of strain that is needed to trigger a plastic event, $\langle\delta\gamma\rangle$, decreases in a power-law manner as a function of the system size $N$ as $\langle\delta\gamma\rangle\sim N^{-\chi}$ with $\chi\approx 2/3$~\cite{Karmakar2010a,Oyama2021}.}
Moreover, many numerical calculations have measured the VDoS in the low frequency limit directly and reported that the non-Debye scaling is indeed universally observed in various  glass systems~\cite{Lerner2016,Mizuno2017,Shimada2018,Kapteijns2018,Wang2019,Richard2020}, although the value of the exponent $D(\omega)\sim \omega^4$ is markedly larger than the mean-field prediction, i.e., $D(\omega)\sim \omega^2$.

The second theoretical prediction is related to the thermodynamic \emph{phase transition}.
When a system in the glass phase crosses a specific border in the parameter space, it experiences full-replica symmetry breakage, and there abruptly emerge infinitesimally different (almost identical) metastable states~\cite{Biroli2016}.
This special (marginally stable) glass phase after the transition is distinguished from normal stable glass states and is called the Gardner phase: in the Gardner phase, even an infinitesimal perturbation can trigger a transition between adjacent almost identical metastable states. 
Such a nature of the Gardner phase is reflected in the infinite hierarchy of metabasins in the energy landscape, and several works have confirmed that the Gardner phase can indeed be observed even in finite physical dimensions in hard sphere systems \cite{Berthier2016a,Jin2018}.
However, the parameter space for the Gardner phase is severely limited, and thus far, no work has detected the Gardner phase in a system with softer potentials, such as the Lennard-Jones (LJ) potential or inverse power-law potential, in physical dimensions~\cite{Scalliet2017,Hicks2018,Seoane2018}.
Thus, the universality of the Gardner aspect of the marginal stability is currently a matter of very active debate~\cite{Berthier2019}, unlike the non-Debye scaling, which has been established as a universal feature even in the physical dimensions.
We stress that to date, most studies~\cite{Berthier2016a,Scalliet2017,Hicks2018,Seoane2018} have focused on the exploration of the Gardner phase via parametrization by thermodynamic variables such as the temperature and the density.

In this article, we numerically demonstrate that the so-called avalanche criticality~\cite{Sethna2001} that LJ glasses experience under external shear exhibits a qualitatively similar characteristic to that of the Gardner phase.
To this end, we first focus on the isotropic samples that are obtained by the minimization of the potential energy from completely random configurations (corresponding to an infinitely fast quench to zero temperature $T=0$ from an infinitely large temperature $T=\infty$).
We measure the statistics of the relative mean squared displacement (MSD) during the initial events that these as-quenched samples first encounter under quasistatic shear.
This observation corresponds to an indirect investigation of the characteristics of unperturbed systems for which previous works have denied the existence of the Gardner phase~\cite{Scalliet2017,Hicks2018,Seoane2018}.
As a result, we did not observe any evidence of criticality that would be expected for the Gardner phase, implying that as-quenched isotropic systems are not in the Gardner phase, consistent with the current understanding~\cite{Scalliet2017,Hicks2018,Seoane2018}.
We next conduct the same measurement for the events in the steady state after the macroscopic yielding.
In the steady state, the probability density functions (PDFs) of the MSD during plastic events exhibit criticality that is consistent with what is observed in the Gardner phase.
Our results suggest that the energy landscape of zero-temperature LJ glasses in the sheared steady state possesses a very similar hierarchical structure to that for the Gardner phase.

\section{Methods}
\subsection{System}
The same LJ glass system as the one introduced in refs.~\cite{Oyama2021,Oyama2021PRL} is employed in this study.
The interparticle potential includes the smoothing polynomial terms, and both the potential and force smoothly go to zero at the cutoff distance $r_{ijc}=1.3\sigma_{ij}$, where $\sigma_{ij}$ is the characteristic interaction range between particles $i$ and $j$.
The system is composed of a $50:50$ mixture of two types of particles that have the same mass and different sizes ($1:1.4$).
We study the response of this system to the external shear.
In particular, we apply the so-called athermal quasistatic (AQS) shear, in which the thermal fluctuations are ignored (zero temperature) and the shear rate is zero~\cite{Kobayashi1980}.
The quasistatic process can be achieved by repeating the exertion of a very small global shearing deformation of $\Delta\gamma$ and the minimization of the total potential energy.
The precise numerical parameters used for the AQS simulation are drawn from ref.~\cite{Oyama2021}~\footnote{Importantly, it has been shown that the statistics of plastic events depend on the strain increment $\Delta\gamma$ per AQS numerical step and that $\Delta\gamma$ should be sufficiently small to obtain a fair evaluation~\cite{Oyama2021}. We use values from ref.~\cite{Oyama2021} that are confirmed to be sufficiently small such that the statistics converge.}.
The FIRE algorithm~\cite{Bitzek2006} is employed for the energy minimization.
The systems are in two dimensions, and the Lees-Edwards periodic boundary conditions~\cite{Allen1987} are set.
The initial configurations are all generated by minimizing the potential energy (corresponding to an infinitely fast quench) from completely random structures (corresponding to infinite temperature).

Amorphous solids are known to experience plastic events, or irreversible rearrangements of constituent particles, under exertion of an external shearing deformation.
In particular, plastic events under an AQS shear can be viewed as transitions between metabasins on the potential energy landscape.
Thus, in this article, we investigate the shape of the potential energy landscape via information on the plastic events under AQS shear.
We note that with a standard shearing protocol, however, we are unable to unambiguously compare the configurations before and after a plastic event even under the AQS condition because the boundary condition is slightly different due to the applied small stepwise strain of $\Delta\gamma$~\footnote{The difference between the configurations before and after a plastic event can be approximately replaced with nonaffine displacements during the event.}.
To rule out such ambiguities introduced by the numerical protocols, we have invented the rewinding method~\cite{Oyama2021}.
In this rewinding method, the strain is \emph{rewinded} by one step (i.e., it is applied in the inverse direction) every time a plastic event is detected (when the stress drops).
This operation allows us to compare configurations of different metabasins under the exact same boundary condition.

\subsection{Relative mean squared displacements}
Our aim here is to study the Gardner-phase-like feature of plastic events under an AQS shear.
As an \emph{order parameter} for the Gardner phase, the PDF of the \emph{overlaps} between different realizations of configurations can be utilized~\cite{Parisi2020}.
As quantitative measures of overlaps, several observables have been proposed~\cite{Parisi2020}.
Among them, we employ the relative MSDs, $\Delta_{AB}\equiv \frac{1}{N}\sum_i^{N}(\boldsymbol{r}_i^A-\boldsymbol{r}_i^B)^2$, because they can be defined without introducing any extra parameters.
Here, $\boldsymbol{r}_i^s$ is the position of particle $i$ in the configuration $s\in {A, B}$, where A and B are used for the states before and after a plastic event, respectively.
{According to the mean-field replica theory, in the Gardner phase, the system shows a \emph{wide} PDF of the MSD (i.e., one that does not converge to a delta peak even in the thermodynamic limit), reflecting the diverging correlation length~\cite{Charbonneau2015,Scalliet2017}.
Since the MSD $\Delta_{AB}$ is defined as an intensive variable, the maximum value of the MSD, which reflects the system-spanning structure, is expected to remain unchanged across different system sizes if the systems are in the Gardner phase (otherwise, we trivially expect $\Delta_{AB}\to 0$ in the thermodynamic limit)~\cite{Parisi2020}.}


\begin{figure}
  \includegraphics[width=\linewidth]{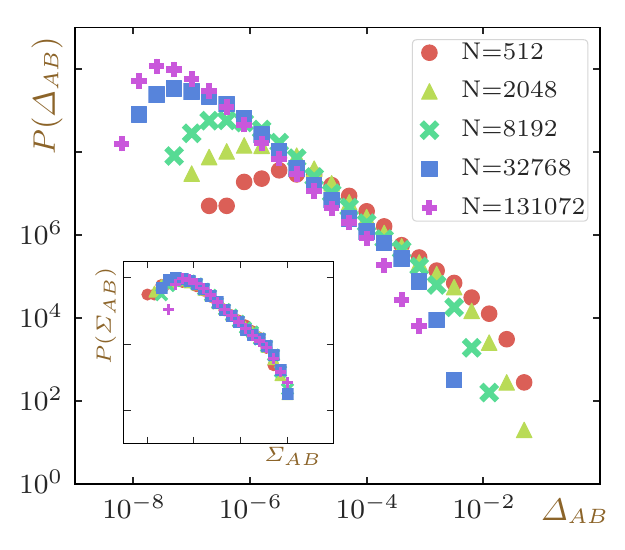}  \caption{
PDFs of the relative MSD during initial events observed in different as-quenched samples.
Different symbols indicate the results for different system sizes, as shown in the legend.
Inset: the corresponding PDFs of the TSD.
    \label{fig:PDF}}
\end{figure}
\section{Results}
\subsection{Initial event ensemble}
We first measure the statistics of the initial events that the as-quenched samples first experience under shear.
These initial events should reflect the features of unperturbed states most strongly.
Different system sizes, i.e., $N=512, 2048, 8192, 32768, 131072$, are investigated in this study, and
we measure the MSDs $\Delta_{AB}$ during the initial events of 4000 independent samples for each system size.
Note that it is known that these initial events do not exhibit the so-called avalanche criticality~\cite{Karmakar2010a,Oyama2021}.

In Fig.~\ref{fig:PDF}, the PDFs of the MSDs during initial events are plotted.
Different symbols stand for different system sizes as shown in the legend.
As evident in Fig.~\ref{fig:PDF}, both the maximum and minimum edges of the PDFs shift toward the smaller $\Delta_{AB}$ side with the increase in $N$.
{This behavior is qualitatively at variance with what we expect for the Gardner phase, where the maximum edge of $P(\Delta_{AB})$ should be constant regardless of the system size, as explained above.}
Rather, the PDFs are consistent with those of the avalanche sizes~\cite{Oyama2021}, indicating the non-system-spanning nature.
This localized (non-system-spanning) tendency can be more explicitly quantified by the total squared displacement (TSD) $\Sigma_{AB}\equiv N\Delta_{AB}$, which carries information regarding the geometrical size of plastic events. 
The PDFs of the TSDs of different system sizes nearly overlap each other without any scaling (see Fig.~\ref{fig:PDF}, inset), as is the case for the PDFs of the avalanche sizes~\cite{Oyama2021}.
Our results imply that the unperturbed systems do not share the same system-spanning vulnerability that is expected for the Gardner phase.
In refs.~\cite{Scalliet2017,Hicks2018,Seoane2018}, the existence of the Gardner phase in physical, finite-dimensional soft-potential systems at rest (that correspond to as-quenched unperturbed samples) was disproved.
Our results are consistent with these studies.

\begin{figure}
  \includegraphics[width=\linewidth]{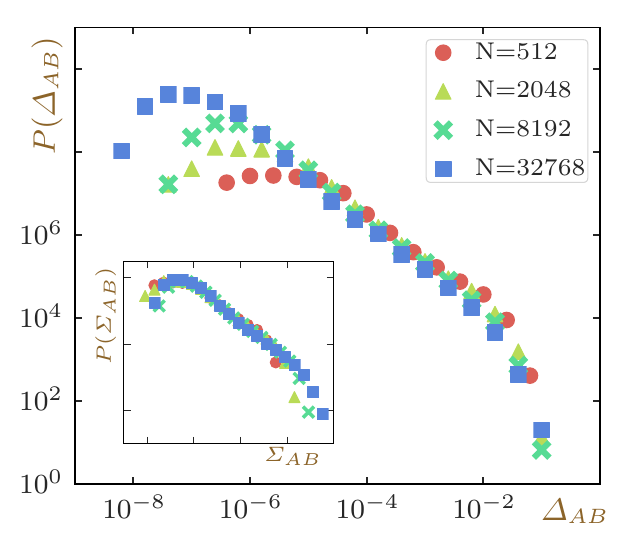}  
\caption{
PDFs of the relative MSD during plastic events in the sheared steady state.
Different symbols indicate the results for different system sizes, as shown in the legend.
Inset: the corresponding PDFs of the TSD.\label{fig:PDF_steady}}
\end{figure}
\subsection{Events in the steady state}
We next measure the same statistics of MSDs during plastic events observed in the steady state.
Since the macroscopic yielding is known to be observed at $\gamma\approx 0.1$~\cite{Ozawa2018}, we can safely treat $\gamma>0.25$ as the steady state, where $\gamma$ is the total accumulated applied strain.
We have gathered 5000 plastic events in the steady state and measured the statistics of MSDs for each system size (for the steady state, we consider $N=512, 2048, 8192, 32768$).
Note that it is known that plastic events in the steady state can form system-spanning avalanches and thus the system exhibits the so-called avalanche criticality, unlike the case of the initial events~\cite{Oyama2021}.

We plot the PDFs of the relative MSD $\Delta_{AB}$ during events in the steady state in Fig.~\ref{fig:PDF_steady}.
The PDFs of the MSD for different system sizes all have broad distributions, and in particular, the edges for the maximum values match well~\footnote{Correspondingly, the maximum edge of the TSD depends linearly on $N$, as shown in Fig.~\ref{fig:PDF_steady} inset.}.
This is exactly what we expect for the Gardner phase.
To put it another way, the avalanche criticality (more specifically, the yielding criticality~\cite{Lin2014PNAS, Ferrero2019PRL}) of plastic events in binary LJ glasses in the steady state has a remarkably similar statistical property to that in the Gardner phase with respect to the spatial structure, although we are still not sure how tightly we can connect these two concepts because of the lack of a theoretical description.
Jin and coworkers~\cite{Jin2018} reported the existence of a shear-induced Gardner transition in a hard sphere system.
The extension of their work to softer potentials would be one promising way to test whether our findings have similar characteristics to theirs.

Since the MSD $\Delta_{AB}$ is a particle-averaged variable, events with the same TSD result in smaller values of $\Delta_{AB}$ in larger systems (compare the plots in the main panel to those in the inset of Fig.~\ref{fig:PDF_steady}. Note that the minimum value of the TSD does not depend on the system size because it represents localized single STZs, the realization of one of which is visualized in Fig.~\ref{fig:vis}(a)).
We note that this characteristic leads to a difference in the widths of $P(\Delta_{AB})$ depending on the system size: the larger the system becomes, the wider $P(\Delta_{AB})$ becomes (the smaller the minimum edge becomes).
Reflecting this difference in the widths, the $P(\Delta_{AB})$ values for different system sizes slightly deviate from each other.
In Appendix~\ref{ap:1}, we demonstrate that they can be collapsed by scaling as $L^\psi P(\Delta_{AB})$ with $\psi\approx 0.5$.
Moreover, in agreement with the PDFs of avalanche sizes~\cite{Oyama2021}, the power-law regime can be seen in a small-value regime, and there is a bump in the large-value regime. 
We mention that the \emph{precursor/mainshock} decomposition proposed in ref.~\cite{Oyama2021} is also valid for PDFs of the MSD, and again, the bump is composed only of mainshocks (see Appendix~\ref{ap:2}).

\subsection{Susceptibilities}
\begin{figure}
  \includegraphics[width=\linewidth]{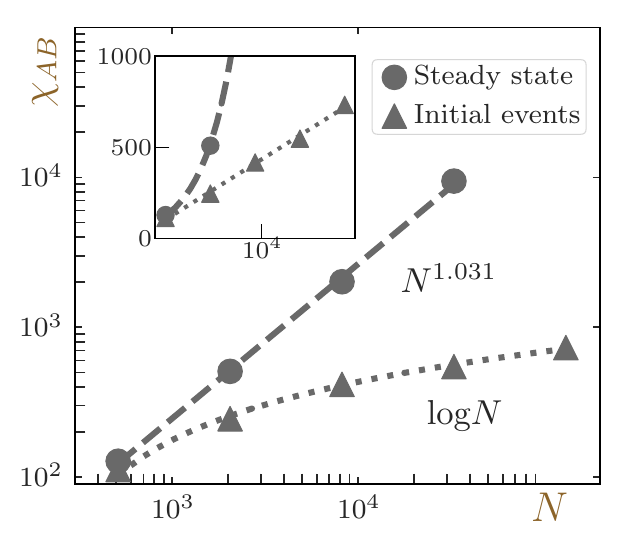}
\caption{
Log-log plot of the susceptibility $\chi_{AB}$ as a function of the system size $N$.
Circles represent the results for the steady state, and triangles, for the initial event ensemble, as shown in the legend.
The inset shows the semi-log plot, with the vertical axis being normal.
Lines represent the fitting results.
    \label{fig:chi_AB}}
\end{figure}

\begin{figure*}
\includegraphics[width=\linewidth]{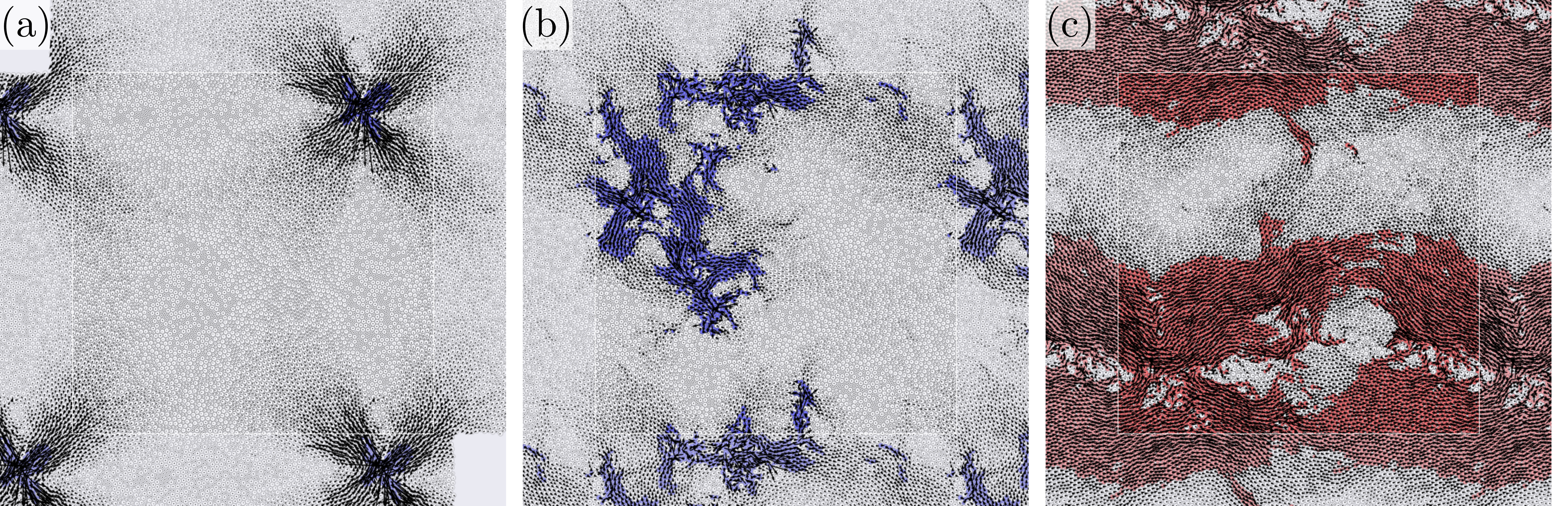}
\caption{
Visualizations of displacement fields during plastic events.
(a) Displacement field during an initial event where only one single STZ is excited.
(b) Displacement field during an initial event when $\Delta_{AB}$ is at its maximum value.
(c) Displacement field during an event in the steady state when $\Delta_{AB}$ is at its maximum value.
All results are from systems with $N=8192$.
Arrows depict the displacement vectors for constituent particles, and their magnitudes are normalized according to the value of the TSD during each event.
Colored particles are \emph{mobile particles} that are determined according to the participation ratio $e=(\sum_i^N d_i^2)^2/(N\sum_i^N d_i^4)$, where $d_i$ is the magnitude of the displacement vector of particle $i$~\cite{Oyama2021}.
The copied images due to the periodic boundary conditions are visualized with slightly lighter colors, and the original cell is indicated by the white square.
    \label{fig:vis}}
\end{figure*}

To further quantify the Gardner-phase-like critical behavior, we next measure the global fluctuations of the relative MSDs.
Obeying ref.~\cite{Scalliet2017}, we employ the following susceptibility $\chi_{\rm AB}$ as the quantitative measure:
\begin{align}
\chi_{\rm AB}=N\cfrac{\langle\Delta^2_{\rm AB}\rangle-\langle\Delta_{\rm AB}\rangle^2}{\langle{\Delta^i_{\rm AB}}^2\rangle-\langle\Delta_{\rm AB}^i\rangle^2},
\end{align}
where $\langle\Delta_{AB}\rangle$ stands for the ensemble averaged relative MSD over distinct plastic events and $\langle \Delta_{AB}^i\rangle =\frac{1}{N}\sum_i^N\langle(\boldsymbol{r}_i^A-\boldsymbol{r}_i^B)^2\rangle$ is the single-particle-based version.
This susceptibility corresponds to the ratio between the collective and microscopic single-particle {configurational changes} and provides the quantitative estimation of the correlation volume.
We show the results for both the initial event ensemble and the events at the steady state in Fig.~\ref{fig:chi_AB}.
Here, $\chi_{\rm AB}$ is plotted as a function of the system size $N$.
In the case of the initial event ensemble, we observe logarithmic divergence of $\chi_{\rm AB}$ as $N$ increases (the logarithmic nature is evident particularly in the inset that is presented in a semi-log manner).
This divergence derives from the long-range nature of the plastic strain in two-dimensional systems~\cite{Picard2004} and does not indicate the system-spanning nature.
Actually, if we visualize the displacement field during events, we can observe that even the event with the largest MSD value does not span the system (Fig.~\ref{fig:vis}(b))~\footnote{We emphasize that the displacement field is still composed of a localized avalanche and is very different from the visualization of an event with a single STZ shown in Fig.~\ref{fig:vis}(a).}.
In other words, the results for the initial event ensemble are again qualitatively different from those expected for the Gardner phase.

On the other hand, the susceptibility of MSDs during steady-state plastic events exhibits a nontrivial power-law divergence as a function of the system size $N$ (evident in the log-log plot in the main panel of Fig.~\ref{fig:chi_AB}).
This seems to correspond to the shear-induced avalanche (yielding) criticality reported in refs.~\cite{Karmakar2010a,Oyama2021} and reflects the fact that STZs tend to form system-spanning avalanches in the steady state (see also the visualization of the event with the largest MSD value shown in Fig.~\ref{fig:vis}(c)).
This power-law divergence of the susceptibility is what we expect for the Gardner phase~\cite{Charbonneau2015,Berthier2016a,Scalliet2017}.
To summarize, from the perspective of the statistics of MSDs, glass systems obtain the Gardner-phase-like feature after a large shear is applied, while they originally lack such characteristics when isotropic.

\section{Discussion}
Here, a question naturally arises: can this qualitative similarity be understood as a shear-induced Gardner transition?
Unfortunately, we cannot provide any absolute conclusion regarding this point at the moment.
Below, we discuss the reasons for this.

First, since our system is athermal, or at zero temperature, our analysis is on the potential energy landscape, not the free energy landscape, which most replica theories are based on.
Moreover, it is still not completely clear whether the energy landscape itself indeed experiences a qualitative change between an unperturbed isotropic state and the steady state.
It may be possible that the energy landscape itself has the Gardner-phase-like feature even in the isotropic state and that the system-spanning nature may become detectable in the steady state just because large deformations cause large interbasin transitions that cannot be activated by small deformations.
Such a interpretation goes along with the claims in refs.~\cite{Parisi2017,Shang2020}.

We also note that we indirectly explored the energy landscape surface via plastic events.
When an avalanche is formed, the system may be able to travel over several metabasins during a single plastic event.
We cannot rule out the possibility that such jumps over multiple metabasins lead to pseudo-Gardner-phase-like criticality, although the real energy landscape does not possess a corresponding structure.

To exclude these possible counterstories and support the argument that the system properties indeed change qualitatively, the measurement of the VDoS can be utilized.
If the Gardner-phase-like features observed in this study mean that the system approaches the situation in the infinite-dimensional mean-field theory due to the exertion of shear, one can expect the exponent of non-Debye scaling to similarly approach the mean-field prediction, $D(\omega)\sim \omega^2$. 
We mention that in a recent article~\cite{Krishnan2021}, the authors reported a strong dependence of the exponent of the non-Debye scaling on the stress ensemble (or equivalently, the strain ensemble) from which the samples are drawn from, even in the vicinity of the isotropic (zero stress) state.

Finally, as another important related work, we mention a prediction of replica theory on the avalanche statistics.
Franz and Spigler discussed the relation between the statistics of avalanche sizes and the energy landscape of the Gardner phase in an amorphous solid system with a genuinely short-range potential in which the jamming criticality also plays a major role~\cite{Franz2017}.
They first formulated the hierarchical structure of the energy landscape in the Gardner phase.
They then treated plastic events under external shear as transitions between metabasins with perturbations induced by shear and showed that, corresponding to the nature of the Gardner phase, such static avalanches are scale-free, and their PDF exhibits power-law behavior.
Their theoretical prediction even provides the quantitative value of the critical exponent for the avalanche size distribution $\tau$, defined as $P(S)\sim S^{-\tau}$, where $S$ stands for the avalanche size and $P(S)$ is its PDF.
Importantly, their prediction argues that the values of $\tau$ are different between systems exactly at the jamming point ($\tau\approx 1.413$) and above jamming ($\tau=1.0$).
Since this theory is based on the replica method, an \emph{equilibrium} statistical mechanics theory, the initial event ensemble that represents the unperturbed system is expected to correspond well to the situation of this theory, although thus far, the plastic events (or avalanches) during a small but finite range of strains have been treated as numerical counterparts~\cite{Franz2017,Shang2020}.
In the present article, we quantified the degree of the similarity between the statistics of the relative MSDs during plastic events under shear and that in the Gardner phase and showed that the presupposition of the full breakage of the replica symmetry in the theory in ref.~~\cite{Franz2017} is not satisfied in the case of the initial event ensemble in our LJ glass system.
This discrepancy is likely to be one of the reasons why the numerical results for the initial event ensemble studied in ref.~\cite{Oyama2021} are inconsistent with the theoretical predictions~\cite{Franz2017}, such as those for criticality and the value of the critical exponent $\tau$.

\section{Conclusions and overview}
In this study, we numerically investigated the statistics of the relative MSDs during plastic events invoked by AQS shear in 2D LJ glasses.
In particular, we considered two distinct ensembles. The first is composed of only the initial events that isotropic as-quenched samples first experience.
The second is composed of the events in the steady state.
In the case of the initial event ensemble, the entire PDF curves of the MSD, $P(\Delta_{AB})$, shift as the system size $N$ increases, and the TSD does not show any major system size dependence.
This result suggests the absence of criticality~\cite{Oyama2021} and is at odds with the behaviors expected for the Gardner phase.
In the case of the ensemble of events in the steady state, on the other hand, the system size dependence of $P(\Delta_{AB})$ qualitatively differs from that for the initial event ensemble: the maximum edges of PDFs of different system sizes remain the same, while the minimum edges become smaller as $N$ increases.
The maximum edges of $P(\Delta_{AB})$ correspond to system-spanning critical events and reflect the fact that the avalanche criticality emerges~\cite{Oyama2021}.
Such an emergence of the criticality is also quantified by the susceptibility defined based on the fluctuations of MSDs, $\chi_{AB}$:
while $\chi_{AB}(N)$ changes in a logarithmic manner for the initial event ensemble, it follows a power-law divergence for steady-state events.
Since all these plastic events can be viewed as transitions between energy metabasins, we can expect that the potential energy landscape of zero-temperature glasses driven into nonequilibrium steady states by AQS shear is qualitatively very similar to that in the Gardner phase.
Again, however, it is important to note that at this stage, we cannot completely rule out the possibility that the as-quenched samples also have the Gardner-phase-like hierarchical energy landscape or, conversely, that the systems in the steady state still do not have such an energy landscape, since we explored the surface of the potential energy landscape indirectly by relying on the information of plastic events under shear, which might have introduced unintentional additional effects.
Nevertheless, we believe that our results here are a good first step toward investigating why the Gardner phase is suppressed in finite-dimensional (isotropic) systems.
In particular, it would be very meaningful to disclose how the system acquires the criticality under shear; we leave this question for future work.

Recently, the similarity between active matter and nonequilibrium systems under shear has been actively discussed~\cite{Tjhung2020a,Morse2021}.
It would be very interesting to investigate a similar emergence of Gardner-phase-like features in various active systems~\cite{Fily2012a,Berthier2017a,Oyama2019c,Tjhung2020a,Morse2021}.
This consideration raises an important question: can the macroscopic yielding be defined in active systems in general~\cite{Morse2021}?

\begin{acknowledgments}

We thank Hajime Yoshino and Harukuni Ikeda for the enlightening
discussions.
This work was financially supported by KAKENHI grants
(nos. 18H05225, 19H01812, 19K14670, 20H01868, 20H00128, 20K14436 and 20J00802) and partially supported by the Asahi Glass Foundation.

\end{acknowledgments}

\appendix
\section{PDFs for MSDs in the steady state}

\subsection{Finite size scaling}\label{ap:1}
\begin{figure}[hb]
\begin{center}
    \includegraphics[width=\linewidth]{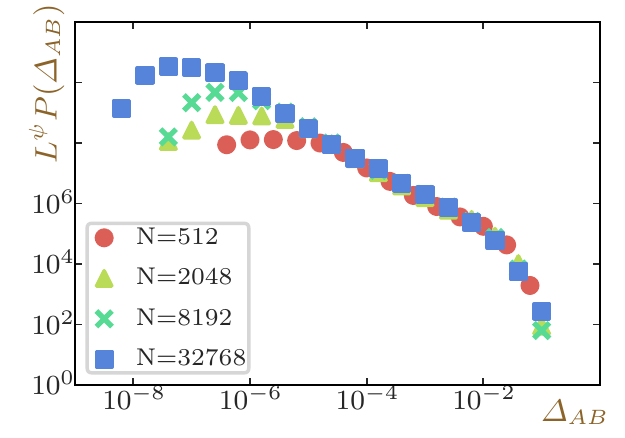}
\caption{\label{fig:SM1}
PDFs of the relative MSD during plastic events in the sheared steady state, scaled by the system linear dimension $L$ with an exponent $\psi$. Different symbols indicate the results for different system sizes, as shown in the legend. }
\end{center}
\end{figure}

As mentioned in the main text, the PDFs of MSDs for different system sizes exhibit deviations from each other, reflecting the difference in their widths.
These curves can be collapsed by scaling as $L^\psi P(\Delta_{AB})$ with $\psi\approx 0.5$, as shown in Fig.~\ref{fig:SM1}.

\subsection{Precursor/mainshock decomposition}\label{ap:2}
\begin{figure}[hb]
\begin{center}
    \includegraphics[width=\linewidth]{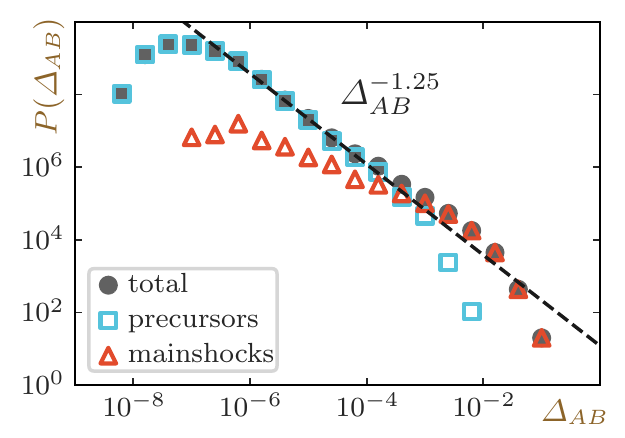}
    \caption{\label{fig:SM2}
      The decomposition of $P(\Delta_{AB})$ into contributions from precursors and mainshocks. The results are for the system with $N=32768$.
      The dashed line depicts the power-law behavior $\Delta_{AB}^{-1.25}$ and is shown as a guide for the eye.
}
\end{center}
\end{figure}

The PDFs of MSDs exhibit bumps in the large-value regime, in accordance with the avalanche size distribution, $P(S)$~\cite{Oyama2021}.
In ref.~\cite{Oyama2021}, it was also shown that $P(S)$ can be decomposed into contributions from \emph{precursors} and \emph{mainshocks} (these are defined based on the global 
trend of the stress strain curves) and that, moreover, the bumps are composed only of mainshocks. 
In Fig.~\ref{fig:SM2}, we demonstrate that $P(\Delta_{AB})$ can also be decomposed into contributions from precursors and mainshocks and that the bump is composed of only mainshocks, similar to $P(S)$~\cite{Oyama2021}.

%
%

%
\end{document}